\documentclass[journal,draftclsnofoot,onecolumn]{IEEEtran}

\usepackage{amssymb}

\usepackage{amsmath}

\usepackage{cite} 

\usepackage[square, comma, sort&compress, numbers]{natbib}

\begin{document}

\newtheorem{theorem}{Theorem}
\newtheorem{proposition}{Proposition}
\newtheorem{lemma}{Lemma}
\newtheorem{definition}{Definition}
\newtheorem{corollary}{Corollary}
\newtheorem{remark}{Remark}
\newtheorem{example}{Example}

\title{Constructions  and necessities of some permutation polynomials
 \thanks{ This work is supported in part by "Funding for scientific research start-up" of Nanjing Tech University.} }

\author{Xiaogang~Liu  
\thanks{
X. Liu is with
College of Computer Science and Technology,
Nanjing Tech University, 
Nanjing City,
Jiangsu Province,
PR China
211800
     e-mail:liuxg0201@163.com. }
}

\maketitle

\begin{abstract}
Permutation polynomials over finite fields   have important applications in many areas of science and engineering such as coding theory, cryptography, combinatorial design, etc.   In this paper, we  construct several new classes of permutation polynomials, and the necessities of some permutation polynomials are studied.  
\end{abstract}


\section{Introduction}\label{secI}

Let $\mathbb{F}_q$ denote the finite field with $q$ elements, and $\mathbb{F}_q^*$ the multiplicative group for a prime power $q$. If a polynomial $f(x)$ over $\mathbb{F}_q$  induces a bijection map from $\mathbb{F}_q$  to itself, it is called a permutation polynomial (PP). If both $f(x)$ and $f(x)+x$ are permutation polynomials over $\mathbb{F}_q$, $f(x)$ is called a complete permutation polynomial. PPs and CPPs   have attracted researchers' much attention for their wide applications in cryptography, coding theory,    and combinational design \cite{A00015}.

Permutation polynomials attract peoples'  interest for their extraordinary properties and algebraic forms. 
Orthomorphisms  map each maximal subgroup of the additive group of   $\mathbb{F}_q$  half into itself and half into its complement, they have a single fixed point, and are the same as CPPs in even characteristic. 
Nonlinear orthomorphisms (or CPPs)   are of cryptographic interest, and   Mittenthal used them for  the design of nonlinear dynamic substitution device \cite{A000M01,A000M02}. Mann introduced CPPs in the construction of orthogonal Latin squares \cite{A000M}.  
PPs have been applied in the Lay–Massey scheme, the block cipher SMS4, the stream cipher Loiss \cite{DL01,DL02,DL03,FF01}, the design of Hash functions, quasigroups, and also in the constructions of some cryptographically strong functions \cite{V01,V02,V03,V04,V05,V06}. 

A monomial $x^n$ permutes  $\mathbb{F}_q$ if $\textup{gcd}(n,q-1)=1$, they are the simplest kind of permutation polynomials. For binomials and trinomials, the permutation properties are not so easy to determine. Carlitz studied permutation binomials in 1962 \cite{A0003c}. In \cite{A0003cw}, Carlitz and Wells found that for $q$ large enough than $d$, the polynomial $f(x)=x(x^{{q-1}\over d}+a)$ might be a permutation polynomial over  $\mathbb{F}_q$. Hou and Lappano studied   permutation binomials of the form $ax+x^{3q-2}, ax+x^{5q-4}$ \cite{A000152,A00012}.   However, only a limited number of constructions are known for PPs.   More recent constructions of PPs can be found in \cite{D01,A000152,A00012,LL01,L02,A00025,W01,W02,W03,W04,W05,W06,W07}.

In this paper, we construct some new classes of permuation polynomials, to some extent they are modifications of some of the PPs  proposed in \cite{X01}. And we continue the work of \cite{X01} to investigate the necessities of two classes of permutation polynomials
presented therein, where  the sufficient conditions are given. To deal with these classes of permutation polynomials, we mainly use  the unit circle of the finite fields, and the algebraic structures of the polynomials.
 Before  coming to our work in Sections \ref{Section II} and \ref{SecIII}, let us first present the following lemmas which might be useful for our study.

\begin{lemma}\label{l01}\cite{Z01}
Let $d,r>0$ with $d\mid q-1$, and let $h(x)\in \mathbb{F}_q[x]$. Then
$f(x)=x^rh(x^{(q-1)/d})$ permutes $\mathbb{F}_q$ if and only if the following two conditions hold:

\begin{enumerate}
\renewcommand{\labelenumi}{$($\mbox{\roman{enumi}}$)$}
\item
$ \textup{gcd}(r,(q-1)/d)=1;$
\item
$x^rh(x)^{(q-1)/d}$ permutes $\mu_{d}, $ where $\mu_{d}$ denotes the $d$-th root of unity in $\mathbb{F}_q$.
 
\end{enumerate}

\end{lemma}

For each element $x$ in the finite field  $\mathbb{F}_{2^{2m}}$, define $\bar{x}=x^{{2^m}}$. The unit circle of  $\mathbb{F}_{2^{2m}}$ is the set
\[
\mathcal{U}=\{\eta\in \mathbb{F}_{2^{2m}}:\eta^{2^m+1}=\eta \bar{\eta}=1\}.
\]

The following lemma can be verified without much difficulty.
\begin{lemma}\label{l02}
 Each nonzero element $x$ in the finite fields  $\mathbb{F}_{2^{2m}} $ has a unique expression of the following form
\[
x=u\lambda,
\]
with $u\in  \mathbb{F}_{2^{m}}^*$ and $\lambda\in \mathcal{U}.$
\end{lemma}
 
%\begin{remark}
%Sometimes we use $\mu_d$ to denote the $d\textup{th}$ roots of unit of a finite field.
%\end{remark}

\begin{lemma}\label{l03}\cite{A000242}
Let $q=2^k$, where $k$ is a positive integer. The quadratic equation $x^2+ux+v$, where $u,v\in \mathbb{F}_q$ and $u\not=0$, has roots in $\mathbb{F}_q$ if and only if $\textup{Tr}_q(v/{u^2})=0$.
\end{lemma}

\section{Constructions of several classes of permutation polynomials over finite fields}\label{Section II}

 In this section, we construct four classes of permutation polynomials over finite fields. To some extent the first three are modifications of permutation polynomials constructed in \cite{X01}, and the fourth class of permutation polynomials comes from a kind of PPs in \cite{L03}.

\subsection{PPs of type $(bx+\delta)^{2^m+1}+ x^{2^m}+cx$ over $\mathbb{F}_{2^{km}}$}

 In  \cite[Proposition 1]{X01},   X. Xu {\sl et al.} proposed a class of permutation polynomials of the form 
$(x^{2^m}+x+\delta)^{s}+bx$ over $\mathbb{F}_{2^{km}}$, with $b\in \mathbb{F}_{2^{m }}^*$ and $\delta$ can be any value. In the following proposition, we consider PPs over   $ \mathbb{F}_{2^{km}}$ by moving the $2^m\textup{th}$ power term out of the bracket, and adding one more constant $c$.

\begin{proposition}
For positive integers $m,n,k$ with $n=km,2\nmid k$. For any $\delta \in \mathbb{F}_{2^n}$, the polynomial
\[
g(x)=(bx+\delta)^{2^m+1}+ x^{2^m}+cx
\]
is a permutation of $\mathbb{F}_{2^n}$ where $b,c \in   \mathbb{F}_{2^n}\backslash \mathbb{F}_{2}$ satisfying $c={b\over b^{2^{2m}}}$.
\end{proposition}

\begin{IEEEproof}
We prove that $g(x)=d$ has at most one solution for any $d\in \mathbb{F}_{2^n}$, which is equivalent to
\begin{equation}\label{e301}
 x^{2^m}+cx+d=(bx+\delta)^{2^m+1}
\end{equation}
has a unique solution.

It can be verified that $\textup{gcd}(2^m+1,2^n-1)=1$ for $n=km$ when  $2\nmid k$. Let $y=bx+\delta$, then $x={y\over b}+{\delta \over b}$. Equation  (\ref{e301}) can be rewritten as
\begin{equation*}\label{e302}
 ({{y\over b}+{\delta \over b}})^{2^m}+c({{y\over b}+{\delta \over b}})+d=y^{2^m+1},
\end{equation*}
which is equivalent to
\begin{equation*}\label{e303}
y^{2^m+1}+{1\over {b^{2^m}}}y^{2^m}+{c \over b}y+{{ \delta^{2^m}}\over {b^{2^m}}} +{{c \delta}\over b} +d=0.
\end{equation*}
That is 
\begin{equation*}\label{e304}
(y^{2^m}+{c \over b})(y+{ 1\over {b^{2^m}}})+{{c }\over {b^{2^m+1}}} +{{\delta^{2^m}}\over {b^{2^m}}} +{{c \delta}\over b} +d=0.
\end{equation*}
So,
\begin{equation}\label{e305}
 (y+{ 1\over {b^{2^m}}})^{2^m+1}={{c }\over {b^{2^m+1}}} +{{\delta^{2^m}}\over {b^{2^m}}} +{{c \delta}\over b} +d.
\end{equation}
by the assumption. 
Now,  $\textup{gcd}(2^m+1,2^n-1)=1$  means that $y^{2^m+1}$ is a permutation of $\mathbb{F}_{2^n}$. Therefore there is a unique $y$ satisfying equation (\ref{e305}).
\end{IEEEproof}

\begin{example}
Let $ m=2,k=3$, then $n=6$. Let  $\delta \in \mathbb{F}_{2^6}$ be any element, $b, c \in \mathbb{F}_{2^6}\backslash{\mathbb{F}_{2}}$, satisfying $c=b^{48}$.  Using Magma, it can be verified that 
\[
g(x)=(bx+ \delta)^{5}+x^{4}+cx
\]
is  a permutation polynomial over $\mathbb{F}_{2^6}$.
\end{example}

\subsection{PPs of type $(x^{2^m}+x+\delta)^{-s}+bx$ over $\mathbb{F}_{2^{2m}}$}
% Xu, X, Feng, X., Zeng, X.

 In \cite[Proposition 7]{X01},   X. Xu {\sl et al.} proposed a class of permutation polynomials of the form 
$(x^{2^m}+x+\delta)^{-s}+bx$ over $\mathbb{F}_{2^{km}}$, with $b\in \mathbb{F}_{2^{m\over 2}}^*$. In the following proposition, we consider PPs over   $ \mathbb{F}_{2^{2m}}$ by changing $2^m+1$ therein to $2^m+2$, and different range of $\delta$.

\begin{proposition}
Let $s,m$ be positive integers satisfying $(2^m+2)(-s) \equiv 2^m-1 \ (\textup{mod} \ 2^{2m}-1$), where $m$ is an odd integer. Let $\delta \in  \mathbb{F}_{2^m}$, then the polynomial
\begin{equation*}
g(x)=(x^{2^m}+x+\delta)^{-s}+bx
\end{equation*}
is a permutation of $\mathbb{F}_{2^{2m}}$, with $b\in \mathbb{F}_{2^{m}}\backslash \mathbb{F}_{2}$. % satisfying ${{ 1}\over {b^{2^m}}}+{1\over b}+1\not= 0$.
\end{proposition}
\begin{IEEEproof}
Since $\textup{gcd}(2^m+2,2^m+1)=1$, and $\textup{gcd}(2^m+2,2^m-1)=\textup{gcd}(3,2^m-1)=1$ for $m$ odd,  
\begin{equation}\label{e201}
\textup{gcd}(2^m+2,2^{2m}-1)=1.
\end{equation}
To prove that $g(x)$ is a permutation polynomial, it is enough to prove that for any $d\in \mathbb{F}_{2^{2m}}$, $g(x)=d$ has a unique solution. That is
\begin{equation*}\label{e202}
(x^{2^m}+x+\delta)^{-s}=bx+d
\end{equation*}
is satisfied by  at most one $x$.
By (\ref{e201}), taking the $(2^m+2)\textup{th}$ power on both sides of the above equation gives the equivalent equation
\begin{equation}\label{e203}
(x^{2^m}+x+\delta)^{2^m-1}=(bx+d)^{2^m+2}.
\end{equation}

First, if there exists a solution $x$ such that
\begin{equation*}\label{e2032}
 x^{2^m}+x+\delta  =0,
\end{equation*}
then $x={d\over b}$, for the right side of equation (\ref{e203}) is also zero. In this case, the above equation   becomes
\begin{equation}\label{e2033}
 {{d^{2^m}}\over {b^{2^m}}}+{d\over b}+\delta  =0.
\end{equation}.

Second,   let us assume that $x^{2^m}+x+\delta\not=0$.
Since taking the $(2^m+1)\textup{th}$ power, the left side of equation (\ref{e203}) is $1$, the right side is in the unit circle $\mathcal{U}$, that is 
\begin{equation*}\label{e204}
(bx+d)^{2^m+2}=\lambda_0
\end{equation*}
for some $\lambda_0\in \mathcal{U}$. But since $\textup{gcd}(2^m+2,2^{ m}+1)=1$,  
\begin{equation*}\label{e205}
 bx+d =\lambda
\end{equation*}
for some $\lambda\in \mathcal{U}$. Thus 
\begin{equation*}\label{e206}
(bx+d)^{2^m+2}=\lambda^{2^m+2}=\lambda =bx+d.
\end{equation*}
And equation (\ref{e203}) can be rewritten as
\begin{equation*}\label{e207}
(x^{2^m}+x+\delta)^{2^m-1}=bx+d.
\end{equation*}
Since $\delta^{2^m}=\delta$, the left side of the above equation becomes
\begin{equation*}\label{e207}
{{x^{2^m}+x+\delta^{2^m}\over {x^{2^m}+x+\delta }}}=1.
\end{equation*}
So, we have $x={{d+1}\over b}$.

Now, the above two situations can be summarized. For every element $b\in \mathbb{F}_{2^{2m}}$, if $b$ satisfies equation (\ref{e2033}), there are two possibilities for the values of $x$ as considered above. But   $x={{d+1}\over b}$ is not the solution. For substituting it into equation (\ref{e203}), the left side becomes
\begin{equation}\label{e208}
 ({{d^{2^m}}\over {b^{2^m}}}+{d\over b}+\delta +{{ 1}\over {b^{2^m}}}+{1\over b})^{2^m-1} =({{ 1}\over {b^{2^m}}}+{1\over b})^{2^m-1}=0 .
\end{equation}  It is not equal to the right side which now becomes $1$.
If $d$ doesn't satisfy equation (\ref{e2033}), and if $x$ is  a solution of equation (\ref{e203}), then $x^{2^m}+x+\delta \not=0$.  The second situation tells us that the  only solution is $x={{d+1}\over b}$.
\end{IEEEproof}

\begin{example}
Set $ m=3,s=6$. Let  $\delta \in \mathbb{F}_{2^3}$ be any element, and $b\in \mathbb{F}_{2^3}\backslash{\mathbb{F}_{2 }}$.  Using Magma, it can be verified that 
\[
g(x)=(x^8+ x+\delta)^{57}+bx
\]
is  a permutation polynomial over $\mathbb{F}_{2^6}$.
\end{example}

\subsection{PPs of type $x^{2^{m+1}}+b^{\prime}x^2+bx$ over $\mathbb{F}_{2^{2m}}$}

 In   \cite[Proposition 8]{X01},  X. Xu {\sl et al.} proposed a class of permutation polynomials over $\mathbb{F}_{2^{4m}}$, of the form 
$(x^{2^m}-x+\delta)^{2^{3m}+2^{m}}+bx$, with $b\in \mathbb{F}_{2^{m}}^*$. In the following proposition, we consider a type of PPs over   $ \mathbb{F}_{2^{2m}}$, of different form and different range of $b$.

\begin{proposition}For the finite field  $\mathbb{F}_{2^{2m}}$, let $b^{\prime}\in \mathcal{U}  $ be in the unit circle, and $b\notin  \mathbb{F}_{2^{ m}}$ satisfying $b^{2(2^m-1)}{b^{\prime} }^3=1$. Then the linearized polynomial 
\[
g(x)=x^{2^{m+1}}+b^{\prime}x^2+bx
\]
is a permutation polynomial of  $\mathbb{F}_{2^{2m}}$.
\end{proposition}

\begin{IEEEproof}
By the assumption,  it can be checked that
\begin{equation}\label{e23}
b^{\prime}\not=b^{1-2^m}.
\end{equation}
Otherwise from  $b^{2(2^m-1)}{b^{\prime} }^3=1$, we have $b^{ 2^m-1} =1$, contradiction with the condition that  $b\notin  \mathbb{F}_{2^{ m}}$.

Since $g(x)$ is a linearzed polynomial, to verify that it is a permutation polynomial, it is necessary to check that
\begin{equation}\label{e01}
g(x)=x^{2^{m+1}}+b^{\prime}x^2+bx=0
\end{equation}
has only the zero solution.
There are two situations to be considered. 

First  assume that $x\in \mathbb{F}_{2^{m}}^*$ is a solution of (\ref{e01}), then 
\begin{equation*}\label{e02}
g(x)=x^{2^{m+1}}+b^{\prime}x^2+bx=x^2+b^{\prime}x^2+bx=0. 
\end{equation*}
That is
\begin{equation*}\label{e03}
 (1+b^{\prime})x+b=0. 
\end{equation*}
If $b^{\prime}=1$, the above equation becomes $b=0$, contradicton. 
So, let us assume that $b^{\prime}\not=1$, then
\begin{equation*}\label{e04}
x={{b}\over{1+b^{\prime}}}. 
\end{equation*}
But we have  $x\in \mathbb{F}_{2^{m}}^*$, that is $x^{2^m}=x$, thus
\begin{equation*}\label{e05}
 {{b^{2^m}}\over{1+{b^{\prime}}^{2^m}}}={{b}\over{1+b^{\prime}}}. 
\end{equation*}
So, 
\begin{equation*}\label{e06}
 {{b^{2^m}b^{\prime}}\over{1+{b^{\prime}} }}={{b}\over{1+b^{\prime}}},
\end{equation*}
which implies that
\begin{equation*}\label{e07}
 b^{\prime}=b^{1-2^{m}},
\end{equation*}
contradiction with equation (\ref{e23}).

Second let us assume that $x\in \mathbb{F}_{2^{2m}}\backslash \mathbb{F}_{2^{m}}$, by Lemma \ref{l02}, we can write 
\begin{equation*}\label{e08}
x=u\lambda
\end{equation*}\label{e08}
  with $u\in  \mathbb{F}_{2^{m}}^*$ and $\lambda\in \mathcal{U}.$ Substituting the above $x$ into equation (\ref{e01}), 
\begin{equation*}\label{e09}
g(x)=x^{2^{m+1}}+b^{\prime}x^2+bx=u^2{1\over {\lambda^2}}+b^{\prime}u^2{\lambda^2}+bu\lambda=0.
\end{equation*}
That is
\begin{equation}\label{e10}
u{1\over {\lambda^2}}+b^{\prime}u{\lambda^2}+b\lambda=u({1\over {\lambda^2}}+b^{\prime}{\lambda^2})+b\lambda=0.
\end{equation}
If $\lambda^4={1\over {b^{\prime}}}= {b^{\prime}}^{2^m}$. The above equation becomes $b\lambda=0$, contradiction. So,    $\lambda^4\not={1\over {b^{\prime}}}$, that is 
\begin{equation}\label{e24}
{1\over {\lambda^2}}+b^{\prime} {\lambda^2}\not=0.
\end{equation}
Then from equation (\ref{e10}), 
\begin{equation*}\label{e11}
 u={{b\lambda}\over b^{\prime}{\lambda^2}+{1\over {\lambda^2}}}.
\end{equation*}
Since  $u\in \mathbb{F}_{2^{m}}^*$, we have  that  $u^{2^m}=u$. The above equation becomes
\begin{equation*}\label{eq12}
\begin{array}{lll}
{{b\lambda}\over b^{\prime}{\lambda^2}+{1\over {\lambda^2}}}&=&{{b^{2^m}\lambda^{2^m}}\over {b^{\prime}}^{2^m}{\lambda^{2^{m+1}}}+{1\over {\lambda^{2^{m+1}}}}}\\
&=&   {{b^{2^m}{1\over {\lambda}}}\over {b^{\prime}}^{2^m}{1\over {\lambda^2}}+{\lambda^2}}.    
 \end{array}
\end{equation*}
That is
\begin{equation*}\label{eq13}
{{b\lambda^3}\over b^{\prime}{\lambda^4}+1}=   {{b^{2^m}{ {\lambda}}}\over {b^{\prime}}^{2^m} +{\lambda^4}}    \Longleftrightarrow {{b\lambda^2}\over b^{\prime}{\lambda^4}+1}=   {{b^{2^m}{  }}\over {b^{\prime}}^{2^m} +{\lambda^4}},
\end{equation*}
which can be rewritten as 
\begin{equation}\label{eq14}
b\lambda^6+b{b^{\prime}}^{2^m}\lambda^2=b^{\prime}b^{2^m}{\lambda^4}+b^{2^m}  \Longleftrightarrow  \lambda^6+b^{\prime}b^{2^m-1}{\lambda^4}+ {b^{\prime}}^{2^m}\lambda^2+b^{2^m-1} =0.
\end{equation}
Let $\lambda_0=\lambda^2$,   equation (\ref{eq14}) can be transformed into
\begin{equation}\label{e15}
  \lambda_0^3+b^{\prime}b^{2^m-1}{\lambda_0^2}+ {b^{\prime}}^{2^m}\lambda_0+b^{2^m-1} =0.
\end{equation}
Take derivative of the above equation
\begin{equation*}\label{eq16}
  \lambda_0^2+  {b^{\prime}}^{2^m } =0.
\end{equation*}
Substituting $\lambda_0^2= {b^{\prime}}^{2^m }$ into equation (\ref{e15})
\begin{equation*}\label{e17}
 {b^{\prime}}^{2^m}\lambda_0+ {b^{\prime}}^{2^m+1}b^{2^m-1} + {b^{\prime}}^{2^m}\lambda_0+b^{2^m-1} =b^{2^m-1}+b^{2^m-1}=0.
\end{equation*}
That is  $\lambda_0= {b^{\prime}}^{2^{m-1} }$ is a double root of (\ref{e15}), which has three roots at most counting multiplicity. But,
\begin{equation*}\label{eq18}
  \lambda^4=\lambda_0^2=  {b^{\prime}}^{2^m }  
\end{equation*}
  contradiction with equation (\ref{e24}).

The third root of (\ref{e15}) is 
\begin{equation*}\label{eq20}
 \lambda_1={b^{2^m-1}\over  {b^{\prime}}^{2^m }  }=b^{2^m-1}b^{\prime}.
\end{equation*}
So
\begin{equation*}\label{eq21}
 \lambda^2= \lambda_1=b^{2^m-1}b^{\prime}  \Longleftrightarrow   \lambda^4=b^{2(2^m-1)}{b^{\prime} }^2.
\end{equation*}
And equation (\ref{e10}) becomes
\begin{equation*}\label{e22}
 u({{1 +b^{\prime}{\lambda^4}}\over {\lambda^2}})+b\lambda=u({{1+b^{2(2^m-1)}{b^{\prime} }^3 }\over {\lambda^2}})+b\lambda=b\lambda=0
\end{equation*}
by assumption, contradiction.
\end{IEEEproof}

\begin{example}
Set $ m=4$. Let  $b^{\prime} \in \mathcal{U}$ be any element of the unit circle in $\mathbb{F}_{2^8}\backslash\mathbb{F}_{2^4}$, $b\in \mathbb{F}_{2^8}\backslash{\mathbb{F}_{2^4}}$ satisfying $b^{30}{b^{\prime} }^3=1$.  Using Magma, it can be verified that 
\[
g(x)=x^{32}+b^{\prime}x^2+bx
\]
is  a permutation polynomial over $\mathbb{F}_{2^8}$.
\end{example}

\subsection{PPs of type $x^r(x^{q-1}+a)$ over $\mathbb{F}_{q^{e}}$}

 In  \cite[Theorem 1]{L03}, K. Li {\sl et al.}  studied a class of permutation polynomials of the form 
$x^r(x^{q-1}+a)$ over $\mathbb{F}_{q^{2}}$,   necessary and sufficient conditions are given. In the following proposition, we consider the same kind of PPs, but over different field  $ \mathbb{F}_{q^{e}}$,  with two particular values of $r$.

\begin{proposition}
Let $\mathbb{F}_{q}$ be the finite field with $q$ elements, then
\[
g(x)=x^r(x^{q-1}+a)
\]
is a permutation polynomial over $\mathbb{F}_{q^e}$ for $r=1, q^{e-1}+q^{e-2}+\cdots+q^2+1$. Here $a\in \mathbb{F}_{q^e}^*$ satisfying $a^{q^{e-1}+q^{e-2}+\cdots+q+1}\not=(-1)^e$, and \textup{gcd}(e-1,q-1)=1. 
\end{proposition}

\begin{IEEEproof}
First, we consider the case $r=1$. Then
\begin{equation*}\label{e601}
g(x)=x^q+ax,
\end{equation*}
which is a linearized polynomial. Since  $a\in \mathbb{F}_{q^e}^*$, and $a^{q^{e-1}+q^{e-2}+\cdots+q+1}\not=(-1)^e$,  it is a PP over $\mathbb{F}_{q^e}$.

 Second, let us consider the case $r=q^{e-1}+q^{e-2}+\cdots+q^2+1$. We found that $d=q^{e-1}+q^{e-2}+\cdots+q+1$ in Lemma \ref{l01}, and $h(x)=x+a$. Thus, $g(x)$ is a permutation polynomial if and only if 
\begin{equation*}\label{e602}
\textup{gcd}(r,q-1)=\textup{gcd}(q^{e-1}+q^{e-2}+\cdots+q^2+1,q-1)=\textup{gcd}(e-1,q-1)=1,
\end{equation*}
and
\begin{equation}\label{e603}
 x^r(x+a)^{q-1}
\end{equation}
permutes $\mu_d$, the $d\textup{th}$ roots of unity in $\mathbb{F}_{q^e}$. Then equation (\ref{e603}) becomes
\begin{equation}\label{e604}
 x^{-q}(x+a)^{q-1} 
\end{equation}
on $\mu_d$. 
Since $\textup{gcd}(-q,q-1)=1$, using Lemma \ref{l01} again, equation (\ref{e604}) permutes $\mu_d$ if and only if
\begin{equation}\label{e605}
 x^{-q}(x^{q-1}+a)
\end{equation}
is a permutation polynomial of $\mathbb{F}_{q^e}$.

Now, equation (\ref{e605}) can be rewritten as 
\begin{equation*}\label{e606}
  x^{-1}+ax^{-q},
\end{equation*}
which is a permutation polynomial of $\mathbb{F}_{q^e}$, since it is linearized if writing $y=x^{-1}$, and $a^{q^{e-1}+q^{e-2}+\cdots+q+1}\not=(-1)^e$.
\end{IEEEproof}

\begin{example}
Let $q=5,e=4,$, and $\omega$ be a primitive root of the finite field $\mathbb{F}_{5^4}$,  then $r=1, q^3+q^2+1=151$. Using Magma, it can be verified that for $a=w^i$, with $1\leq i\leq 623, i\not=0 \ \textup{mod}\ 4$, 
\[
g(x)=x^{r}(x^{4}+a)
\]
is a permutation polynomial over $\mathbb{F}_{5^4}$.

\end{example}

\section{Necessities of two kinds of permutation polynomials}\label{SecIII}

In this section, we investigate the necessities of two classes of permutation polynomials studied in   \cite{X01}, where the
sufficient conditions are given.

\subsection{PPs of type $(x^{2^m}+x+\delta)^{2^{2m-1}+2^{m-1}}+bx$ over $\mathbb{F}_{2^{2m}}$}

 In   \cite[Proposition 10]{X01}, X. Xu {\sl et al.} proposed a class of permutation polynomials of the form 
$(x^{2^m}+x+\delta)^{2^{2m-1}+2^{m-1}}+bx$ over $\mathbb{F}_{2^{2m}}$,   and sufficient conditions are given. In the following proposition, we consider the same kind of PPs,  but its necessary conditions.

\begin{proposition}
For a positive integer $m$ and a fixed $\delta \in \mathbb{F}_{2^{2m}}$ with $\textup{Tr}_m^{2m}(\delta)\not= 0$, let
\[
g(x)=(x^{2^m}+x+\delta)^{2^{2m-1}+2^{m-1}}+bx
\]
where $b\in \mathbb{F}_{2^{2m}}$. When $b\notin \mathbb{F}_{2^{m}}$,  $g(x)$ is permutation polynomial if and only if $b+b^m=b^{2^m+1}$.
\end{proposition}

\begin{IEEEproof}
As pointed out at the beginning of this subsection,   \cite[Proposition 10]{X01} gives the sufficiency verification. Now let us consider the necessity.

Assume that $g(x)$ is a permuatation polynomial. Then for every $d\in \mathbb{F}_{2^{2m}}$, $g(x)=d$ has a unique solution. That is 
\begin{equation*}\label{e401}
(x^{2^m}+x+\delta )^{2^{2m-1}+2^{m-1}}+ bx=d
\end{equation*}
has at most one possibile root in $ \mathbb{F}_{2^{2m}}$. Squaring both sides of the above equation, we get the following equivalent equation
\begin{equation*}\label{e402}
(x^{2m}+x+\delta )^{2^{2m }+2^{m }}+ b^2x^2=d^2.
\end{equation*}
That is
\begin{equation*}\label{e403}
(x^{2^m}+x+\delta )(x^{2^m}+x+\delta^{2^m} )= b^2x^2+d^2, 
\end{equation*}
which can be transformed into
\begin{equation*}\label{e404}
(x^{2^m}+x )^2+( \delta+\delta^{2^m} )(x^{2^m}+x)+\delta^{2^m+1}= b^2x^2+d^2. 
\end{equation*}
Which implies that
\begin{equation}\label{e405}
x^{2^{m+1}}+ ( \delta+\delta^{2^m} )x^{2^m} +(b^2+1)x^2+ ( \delta+\delta^{2^m} )x  +\delta^{2^m+1}+d^2=0
\end{equation}
has a unique solution in $ \mathbb{F}_{2^{2m}}$.

 Then for  $x_1\not= x_2 \in \mathbb{F}_{2^{2m}}$ with $x_1$ a solution of equation (\ref{e405}), the following equation 
\begin{equation*}\label{e406}
x_2^{2^{m+1}}+ ( \delta+\delta^{2^m} )x_2^{2^m} +(b^2+1)x_2^2+ ( \delta+\delta^{2^m} )x_2  +\delta^{2^m+1}+d^2=0
\end{equation*}
can not hold. Adding the above two equations,  
\begin{equation}\label{e407}
(x_1+x_2)^{2^{m+1}}+ ( \delta+\delta^{2^m} )(x_1+x_2)^{2^m} +(b^2+1)(x_1+x_2)^2+ ( \delta+\delta^{2^m} )(x_1+x_2)  =0
\end{equation}
does not hold for any $x_2$ different from $x_1$. Now let $y=x_1+x_2$. With $x_1$ fixed  and $x_2$ varying, $y$ can be any nonzero element of the finite field $ \mathbb{F}_{2^{2m}}$. So,  
\begin{equation}\label{e408}
y^{2^{m+1}}+ ( \delta+\delta^{2^m} )y^{2^m} +(b^2+1)y^2+ ( \delta+\delta^{2^m} )y  =0
\end{equation}
has only the solution zero in $ \mathbb{F}_{2^{2m}}$, this is from the assumption that $g(x)$ is a permutation polynomial. 

If  a nonzero solution  $y\in \mathbb{F}_{2^{2m}} $ of equation (\ref{e408}) exists. Taking the $2^{m}\textup{th}$ power,
\begin{equation}\label{e409}
y^{2 }+ ( \delta+\delta^{2^m} )y +(b^{2^{m+1}}+1)y^{2^{m+1}}+ ( \delta+\delta^{2^m} )y^{2^m}  =0.
\end{equation}
Adding equations (\ref{e408}) and (\ref{e409}),  
\begin{equation*}\label{e410}
 b^{2^{m+1}}y^{2^{m+1}}+b^2y^2   =(b^2y^2 )^{2^m}+(b^2y^2 )=0.
\end{equation*}
Thus 
\begin{equation*}\label{e411}
 (by )^{2^m}+(by )=0, 
\end{equation*}
that is $by$ lies in the field   $ \mathbb{F}_{2^{m}}$. By Lemma \ref{l02}, we can write
\begin{equation}\label{e412}
b={c_0\over \lambda_0 }
\end{equation}
for some fixed $c_0\in  \mathbb{F}_{2^{m}}\backslash{\{0\}}$, and $\lambda_0 \in \mathcal{U}$ the unit circle. If $y$ is written in the following form
\[
y=c\lambda
\]
for $c\in \mathbb{F}_{2^m}$ and $\lambda \in \mathcal{U}$. Since $by\in \mathbb{F}_{2^{m}}$, we must have $\lambda=\lambda_0$. That is
\begin{equation}\label{e413}
y=c\lambda_0
\end{equation}
for some $c\in \mathbb{F}_{2^m}\backslash{\{0\}}$.

Substituting (\ref{e412}) and (\ref{e413}) into equation (\ref{e408})  
\begin{equation*}\label{e414}
 {c^2\over \lambda_0^2}+( \delta+\delta^{2^m} ){c \over \lambda_0 } +( {c_0^2\over \lambda_0^2}+1)c^2\lambda_0^2+ ( \delta+\delta^{2^m} )c\lambda_0 =0.
\end{equation*}
 Dividing $c$ on both sides of the above equation 
\begin{equation*}\label{e415}
 {c \over \lambda_0^2}+( \delta+\delta^{2^m} ){1\over \lambda_0 } +( {c_0^2\over \lambda_0^2}+1)c \lambda_0^2+ ( \delta+\delta^{2^m} ) \lambda_0 =0,
\end{equation*}
which can be transformed into
\begin{equation*}\label{e416}
 ( {c_0^2 }+\lambda_0^2+{1\over \lambda_0^2})c =( \delta+\delta^{2^m} ){1\over \lambda_0 } + ( \delta+\delta^{2^m} ) \lambda_0.  
\end{equation*}

By our assumption, equation (\ref{e408}) has no nonzero solution,  then
\begin{equation*}\label{e417}
  {c_0^2 }+\lambda_0^2+{1\over \lambda_0^2}=0,
\end{equation*}
which is equivalent to 
\begin{equation*}\label{e418}
  {c_0  }=\lambda_0 +{1\over \lambda_0 } 
\end{equation*}
for some $\lambda_0\not=1$ in $\mathcal{U}$, because $b\notin \mathbb{F}_{2^{m}}$.
By equation (\ref{e412}),  
\begin{equation*}\label{e419}
b=1+{1\over \lambda_0^2 }.
\end{equation*}
That is,
\begin{equation*}\label{e420}
b=1+{1\over \lambda }
\end{equation*}
for some $\lambda\in \mathcal{U}\backslash{\{1\}}$.
So, we have
\begin{equation*}\label{e421}
b^{2^m+1}=(1+{1\over \lambda })(1+{  \lambda })={1\over \lambda }+\lambda.
\end{equation*}
And 
\begin{equation*}\label{e422}
b^{2^m  }+b=(1+{  \lambda })+(1+{1\over \lambda })={1\over \lambda }+\lambda.
\end{equation*}
Which implies that 
\begin{equation*}\label{e422}
b^{2^m  }+b=b^{2^m+1},
\end{equation*}
that is the necessity of our proposition.
\end{IEEEproof}

\begin{example}
Let $ m=4$,   $\delta \in \mathbb{F}_{2^8}$ with $\textup{Tr}_m^{2m}(\delta)\not=0$. Using Magma, it can be verified that for $b\in \mathbb{F}_{2^8}\backslash{\mathbb{F}_{2^4}}$, 
\[
g(x)=(x^{16}+x+\delta)^{136}+bx
\]
is not a permutation polynomial over $\mathbb{F}_{2^8}$ when $b^{16}+b\not=b^{17}$.
\end{example}

\subsection{PPs of type $(x^2+x+\delta)^{2^{2k-1}-2^{k-1}}+bx$ over $\mathbb{F}_{2^{2k}}$}

 In   \cite[Proposition 6]{X01}, X. Xu {\sl et al.} proposed a class of permutation polynomials of the form 
$(x^{2 }+x+\delta)^{2^{2k-1}-2^{k-1}}+bx$ over $\mathbb{F}_{2^{2k}}$,   and sufficient conditions are given. In the following proposition, we consider the same kind of PPs,  but its necessary conditions.

\begin{proposition}
For nonnegative integers $n,k$ with $n=2k, k>1$, let $\delta \in \mathbb{F}_{2^n}$ with $\textup{Tr}_1^n(\delta)=1$. Then the polynomial
\[
g(x)=(x^2+x+\delta)^{2^{2k-1}-2^{k-1}}+bx
\]
is a permutation of $\mathbb{F}_{2^n}$ if and only if $b\in \mathbb{F}_{2^k}\backslash{\{0\}}$.
\end{proposition}

\begin{IEEEproof}
The sufficiency is given in   \cite[Proposition 6]{X01}. In the following we only consider the necessity. 

Assume that $b\notin \mathbb{F}_{2^k}$, and $g(x)$ is a PP.

Since  $\textup{Tr}_1^n(\delta)=1$, $x^2+x+\delta$ is always nonzero by Lemma \ref{l03}. For any $d\in \mathbb{F}_{2^n}$, the following equation
\begin{equation*}\label{e501}
(x^2+x+\delta)^{2^{2k-1}-2^{k-1}}+bx=d
\end{equation*}
has only one solution, which  can be transformed into
 \begin{equation}\label{e504}
(x^2+x+\delta)^{2^{2k-1}-2^{k-1}}=bx+d.
\end{equation}
Taking the $(2^k+1)\textup{th}$ power on both sides of the above equation
\begin{equation*}\label{e502}
1=(bx+d)^{2^k+1}.
\end{equation*}
So, $bx+d=\lambda$, that is
\begin{equation}\label{e506}
x={{\lambda+d}\over b}
\end{equation}
for some element $\lambda$ in the unit circle $\mathcal{U}$.
Squaring both sides of equation (\ref{e504})  
\begin{equation}\label{e503}
(x^2+x+\delta)^{1-2^{k }}= (bx+d)^2.
\end{equation}
That is
\begin{equation*}\label{e505}
{{x^2+x+\delta}\over {x^{2^{k+1}}+x^{2^k}+\delta^{2^k}} } = \lambda^2,
\end{equation*}
which is equivalent to 
\begin{equation*}\label{e507}
{{x^2+x+\delta}   }= \lambda^2(x^{2^{k+1}}+x^{2^k}+\delta^{2^k} ).
\end{equation*}
Substituting  (\ref{e506}) into the above equation,  
\begin{equation}\label{e508}
({1\over b^{2}}+{d^{2^{k+1}}\over b^{2^{k+1}}}+{d^{2^{k }}\over b^{2^{k }}}+\delta^{2^k})\lambda^2
+({1\over b}+{1\over b^{2^k}})\lambda
+{{d^2\over b^2}+{d\over b}+{1\over b^{2^{k+1}}}}+\delta=0.
\end{equation}
We can choose $d$ such that
\begin{equation*}\label{e509}
 {1\over b^{2}}+{d^{2^{k+1}}\over b^{2^{k+1}}}+{d^{2^{k }}\over b^{2^{k }}}+\delta^{2^k}\not=0.
\end{equation*}
Since $g(x)$ is a permutation polynomial, there must exists $\lambda_1$ in the unit circle $\mathcal{U}$, satisfying equation (\ref{e508}). Then
\begin{equation*}\label{e510}
x_1={{\lambda_1+d}\over b}
\end{equation*}
satisfies equation (\ref{e503}), and in fact $g(x)=d$, since they are equivalent. Equation (\ref{e508}) can be transformed into
\begin{equation*}\label{e511}
\lambda^2
+{{{1\over b}+{1\over b^{2^k}}}\over {{1\over b^{2}}+{d^{2^{k+1}}\over b^{2^{k+1}}}+{d^{2^{k }}\over b^{2^{k }}}+\delta^{2^k}}}\lambda
+{ ({{1\over b^{2}}+{d^{2^{k+1}}\over b^{2^{k+1}}}+{d^{2^{k }}\over b^{2^{k }}}+\delta^{2^k}})^{2^k}\over {{1\over b^{2}}+{d^{2^{k+1}}\over b^{2^{k+1}}}+{d^{2^{k }}\over b^{2^{k }}}+\delta^{2^k}}}=0.
\end{equation*}
Since ${ ({{1\over b^{2}}+{d^{2^{k+1}}\over b^{2^{k+1}}}+{d^{2^{k }}\over b^{2^{k }}}+\delta^{2^k}})^{2^k}\over {{1\over b^{2}}+{d^{2^{k+1}}\over b^{2^{k+1}}}+{d^{2^{k }}\over b^{2^{k }}}+\delta^{2^k}}} $ is in the unit circle, the other root of equation (\ref{e508}), which we denote $\lambda_2$, is also in the unit circle $\mathcal{U}$. 
And 
\[
\lambda_1\not=\lambda_2,
\]
since $\lambda_1+\lambda_2={1\over b}+{1\over b^{2^k}}\not=0$ for $b\notin \mathbb{F}_{2^k}$.

Now, for equation (\ref{e506}), set
 \begin{equation*}\label{e512}
x_2={{\lambda_2+d}\over b}.
\end{equation*}
Then $x_2$ satisfies equation (\ref{e503}) also, that is $g(x)=d$ has two solutions $x_1,x_2$ for such $d$, contradiction. 
\end{IEEEproof}

\begin{example}
Set $k=4$, then $n=8$. Let $\delta \in \mathbb{F}_{2^n}$ with $\textup{Tr}(\delta)=1$. Using Magma, it can be verified that for $b\in \mathbb{F}_{2^n}\backslash{\mathbb{F}_{2^k}}$, 
\[
g(x)=(x^2+x+\delta)^{120}+bx
\]
is not a permutation polynomial over $\mathbb{F}_{2^8}$.

\end{example}

\section{Conclusion}

In this paper, we construct some classes of permutation polynomials over finite fields, which are modifications of known permutation polynomials recently studied. We also investigate the necessities of   permutation properties of the polynomials  studied in \cite{X01}, where the sufficient conditions are given.

 \section*{Acknowledgment}

 The author would like to thank the anonymous referees for helpful suggestions and comments.

%.............

\end{document}